\documentstyle[psfig,rotate]{mn}
\input{psfig}
\newif\ifAMStwofonts
\ifoldfss

  \ifCUPmtlplainloaded \else
    \NewTextAlphabet{textbfit} {cmbxti10} {}
    \NewTextAlphabet{textbfss} {cmssbx10} {}
    \NewMathAlphabet{mathbfit} {cmbxti10} {}
    \NewMathAlphabet{mathbfss} {cmssbx10} {}
  \fi
  \ifAMStwofonts
    \ifCUPmtlplainloaded \else
      \NewSymbolFont{upmath} {eurm10}
      \NewSymbolFont{AMSa} {msam10}
      \NewMathSymbol{\upi}     {0}{upmath}{19}
      \NewMathSymbol{\umu}     {0}{upmath}{16}
      \NewMathSymbol{\upartial}{0}{upmath}{40}
      \NewMathSymbol{\leqslant}{3}{AMSa}{36}
      \NewMathSymbol{\geqslant}{3}{AMSa}{3E}

       \let\le=\leqslant
       
    \fi
  \fi
\fi 
\ifnfssone
  \newmathalphabet{\mathit}
  \addtoversion{normal}{\mathit}{cmr}{m}{it}
  \addtoversion{bold}{\mathit}{cmr}{bx}{it}

  \newmathalphabet{\mathbfit} 
  \addtoversion{normal}{\mathbfit}{cmr}{bx}{it}
  \addtoversion{bold}{\mathbfit}{cmr}{bx}{it}
  \newmathalphabet{\mathbfss} 
  \addtoversion{normal}{\mathbfss}{cmss}{bx}{n}
  \addtoversion{bold}{\mathbfss}{cmss}{bx}{n}
  \ifAMStwofonts
    \ifCUPmtlplainloaded \else
      \UseAMStwoboldmath
      \makeatletter
      \new@mathgroup\upmath@group
      \define@mathgroup\mv@normal\upmath@group{eur}{m}{n}
      \define@mathgroup\mv@bold\upmath@group{eur}{b}{n}
      \edef\UPM{\hexnumber\upmath@group}
      \new@mathgroup\amsa@group
      \define@mathgroup\mv@normal\amsa@group{msa}{m}{n}
      \define@mathgroup\mv@bold\amsa@group{msa}{m}{n}
      \edef\AMSa{\hexnumber\amsa@group}
      \makeatother
      \mathchardef\upi="0\UPM19
      \mathchardef\umu="0\UPM16
      \mathchardef\upartial="0\UPM40
      \mathchardef\leqslant="3\AMSa36
      \mathchardef\geqslant="3\AMSa3E

       \let\le=\leqslant
       
    \fi
  \fi
\fi 
\ifnfsstwo

  \DeclareMathAlphabet{\mathbfit}{OT1}{cmr}{bx}{it}
  \SetMathAlphabet\mathbfit{bold}{OT1}{cmr}{bx}{it}
  \DeclareMathAlphabet{\mathbfss}{OT1}{cmss}{bx}{n}
  \SetMathAlphabet\mathbfss{bold}{OT1}{cmss}{bx}{n}
  \ifAMStwofonts
    \ifCUPmtlplainloaded \else
      \DeclareSymbolFont{UPM}{U}{eur}{m}{n}
      \SetSymbolFont{UPM}{bold}{U}{eur}{b}{n}
      \DeclareSymbolFont{AMSa}{U}{msa}{m}{n}
      \DeclareMathSymbol{\upi}{0}{UPM}{"19}
      \DeclareMathSymbol{\umu}{0}{UPM}{"16}
      \DeclareMathSymbol{\upartial}{0}{UPM}{"40}
      \DeclareMathSymbol{\leqslant}{3}{AMSa}{"36}
      \DeclareMathSymbol{\geqslant}{3}{AMSa}{"3E}

       \let\le=\leqslant
       
    \fi
  \fi
\fi 
\ifCUPmtlplainloaded \else
  \ifAMStwofonts \else 
    \def\upi{\pi}
    \def\umu{\mu}
    \def\upartial{\partial}
  \fi
\fi
\title{Testing Cold Dark Matter Models Using Hubble Flow Variations}
\author[Xiangdong Shi]
   {Xiangdong Shi\\
   Department of Physics, University of California, La Jolla, CA 92093, USA}
\date{Accepted 
      Received 
      in original form}
\pubyear{1997}
\begin{document}
\maketitle
\begin{abstract}
\baselineskip=20pt
{\it COBE}-normalized flat (matter plus cosmological constant) and
open Cold Dark Matter (CDM) models are tested by comparing
their expected Hubble flow variations and the observed variations in
a Type Ia supernova sample and a Tully Fisher cluster sample.
The test provides a probe of the CDM power spectrum on scales of
$0.02h$ Mpc$^{-1}\la k\la 0.2h$ Mpc$^{-1}$, free of the bias factor $b$.
The results favor a low matter content universe, or a flat matter-dominated
universe with a very low Hubble constant and/or a very small spectral
index $n_{\rm ps}$, with the best fits having $\Omega_0\sim 0.3$ to 0.4.
The test is found to be more discriminative to the
open CDM models than to the flat CDM models.
For example, the test results are found to be compatible with those
from the X-ray cluster abundance measurements at smaller length scales,
and consistent with the galaxy and cluster correlation analysis of
Peacock and Dodds (1994) at similar length scales, if our universe is flat;
but the results are marginally incompatible with the X-ray cluster abundance
measurements if our universe is open.  The open CDM results are consistent
with that of Peacock and Dodds only if the matter density of the universe
is less than about 60\% of the critical density.
The shortcoming of the test is discussed, so are ways to minimize it.
\end{abstract}
\baselineskip=20pt
\section{Introduction}

In a previous paper on Hubble flow variations (Shi 1997a),
I calculated the limit on the power spectrum
shape parameter $\Gamma$ for flat CDM models.
In this paper I extend the test of Hubble flow variations to
open CDM models, and do a likelihood analysis of both type of models.
Furthermore, I will compare the results to those
from other methods that test the CDM power spectrum.
Finally, I will discuss the major shortcoming of this test, and
possible improvements that can be made to minimize it.

Before going into details, I would like to point out first that the method
outlined here is different from that of Jaffe and Kaiser (1995).
In Jaffe and Kaiser (1995), multi-mode deviations from a pure Hubble expansion,
including the bulk motion and shears were investigated for
samples as a whole.  In the method presented here,
the variations of the Hubble expansion, corresponding to the isotropic
component of Jaffe and Kaiser's multi-mode deviation, are investigated
within samples for every subsample (excluding those with too small sizes
such that non-linearity becomes an issue).
Therefore the Hubble flow variation method is
sensitive to the scale dependence and
the shape of the variations, while the method of Jaffe and Kaiser is not.
The two methods are however based on the same theoretical premise:
density fluctuations give rise to peculiar velocities through gravity,
and therefore from the amplitude and direction of peculiar
velocity fields one can infer the underlying density fluctuations.
It will be interesting to combine the two methods, by testing the
variations of all modes of peculiar velocities,
though it will be significantly more cpu intensive.

The formalism of testing models using Hubble flow variations has been
reviewed in Shi (1997a,b).  Here a summarization suffices.
The deviation from a global Hubble expansion rate $H_0$, $\delta H$
($=H_{\rm Local}-H_0$), of a sample is
\begin{equation}
\delta H=B^{-1}\sum_q{S_qr_q-U_ir_q\hat r_q^i\over\sigma_q^2},
\label{pechub}
\end{equation}
where its bulk motion is
\begin{equation}
U_i=(A-RB^{-1})^{-1}_{ij}\Bigl(\sum_q{S_q{\hat r}_q^j\over\sigma_q^2}-
    B^{-1}\sum_q\sum_{q^\prime}{S_qr_qr_{q^\prime}{\hat r}_{q^\prime}^j\over
    \sigma_q^2\sigma_{q^\prime}^2}\Bigr),
\label{bulkmotion}
\end{equation}
and
\begin{equation}
A_{ij}=\sum_q{\hat r_q^i\hat r_q^j\over\sigma_q^2},\,\,
R_{ij}=\sum_q\sum_{q^\prime}{r_q\hat r_q^ir_{q^\prime}
\hat r_{q^\prime}^j\over\sigma_q^2\sigma_{q^\prime}^2},\,\,
B=\sum_q{r_q^2\over\sigma_q^2}.
\label{Aij}
\end{equation}
In the equations, ${\bf r}_q$ is the position of object $q$ in the sample (with
earth at the origin), and $S_q$ ($=cz_q-H_0r_q$) is its estimated line-of-sight
peculiar velocity with an uncertainty $\sigma_q$. Spatial indices $i,j$ run
from $1,2,3$, and identical indices indicate summation.

Formally $\delta H$ can always be expressed in the form
\begin{equation}
\delta H=\int {\rm d}^3r\,\tilde{W}({\bf r})\,S({\bf r}),
\label{pechub1}
\end{equation}
where $\tilde{W}({\bf r})$ is the window function of the $\delta H$
measurement, and $S({\bf r})$ is the estimated line-of-sight
peculiar velocity field.  It is not hard to see from
eqs.~(\ref{pechub}),~(\ref{bulkmotion}) and~(\ref{Aij}) that
\begin{eqnarray}
\tilde{W}({\bf r})&=&
    B^{-1}\Bigl\{\sum_q {r_q\over\sigma_q^2}\delta ({\bf r}-{\bf r}_q)-
    \nonumber\\
&&    (A-RB^{-1})^{-1}_{jl}\Bigl[\sum_q{\hat r_q^j\over\sigma_q^2}
    \delta ({\bf r}-{\bf r}_q)-
    \nonumber\\
&&   B^{-1}\sum_q\sum_{q^\prime}{r_qr_{q^\prime}^j\over
    \sigma_q^2\sigma_{q^\prime}^2}\delta ({\bf r}-{\bf r}_q)\Bigr]
    \sum_{q^{\prime\prime}}{r_{q^{\prime\prime}}^l
    \over\sigma_{q^{\prime\prime}}^2}\Bigr\}.
\end{eqnarray}
If most of the uncertainty in $S_q$ comes from the uncertainty
in measuring the distance $r_q$, which is true at scales beyond
$\sim 5000$ km/sec, then $\sigma_q\propto r_q$.  As a result,
the window function $\tilde{W}$ scales linearly with $H_0$.

The estimated line-of-sight peculiar velocity of object $q$, $S_q$,
is related to its true peculiar velocity ${\bf v}({\bf r_q})$ by
\begin{equation}
S_q=v_i({\bf r}_q)\hat r^i_q+\epsilon_q,
\end{equation}
where $\epsilon_q$ is the uncertainty of the estimate with the
standard deviation of $\sigma_q$.  Therefore,
$\delta H$ can be broken into two parts: the true deviation
$\delta H^{(v)}$ and the noise $\delta H^{(\epsilon)}$.
Eq.~(\ref{pechub1}) then becomes
\begin{equation}
\delta H=\delta H^{(v)}+\delta H^{(\epsilon)}
=\int {\rm d}^3r\,W^i({\bf r})\,v_i({\bf r})\,+\,
\int {\rm d}^3r\,\tilde{W}({\bf r})\,\epsilon,
\end{equation}
where $W^i({\bf r})=\tilde{W}({\bf r})\,\hat r^i$.  The variance of
$\delta H^{(v)}$ depends on the density power spectrum of
our universe $P(k)$, the matter content of our universe $\Omega_0$
(the matter density divided by the critical density), and the global
Hubble constant $H_0$, in the following way (Shi 1997a,b):
\begin{equation}
\bigl\langle\bigl(\delta H^{(v)}\bigr)^2\bigr\rangle
=H_0^2\,\Omega_0^{1.2}\int {\rm d}^3k
\bigl\vert W^i({\bf k})\hat k_i\bigr\vert ^2{P(k)\over k^2},
\label{variance1}
\end{equation}
where the Fourier transform of
$W^i({\bf r})$, $W^i({\bf k})$, is
\begin{eqnarray}
W^i({\bf k})&=&{B^{-1}\over (2\pi)^{3/2}}
   \Bigl[\sum_q {r_q^i\over\sigma_q^2}e^{i{\bf k}\cdot{\bf r}_q}\nonumber\\
&-&(A-RB^{-1})^{-1}_{jl}\Bigl(\sum_q{\hat r_q^i\hat r_q^j\over\sigma_q^2}
   e^{i{\bf k}\cdot{\bf r}_q}\nonumber\\
&-&B^{-1}\sum_q\sum_{q^\prime}{r_q^ir_{q^\prime}^j\over
   \sigma_q^2\sigma_{q^\prime}^2}e^{i{\bf k}\cdot{\bf r}_q}\Bigr)
   \sum_{q^{\prime\prime}}{r_{q^{\prime\prime}}^l
   \over\sigma_{q^{\prime\prime}}^2}\Bigr].
\label{Wk}
\end{eqnarray}
The variance of $\delta H^{(\epsilon)}$ depends on
the sample measures and $H_0$ in the following form:
\begin{equation}
\bigl\langle\bigl(\delta H^{(\epsilon)}\bigr)^2\bigr\rangle
=H_0^2\bigl(B^{-1}+B^{-2}(A-RB^{-1})^{-1}_{il}R_{il}\bigr).
\label{variance2}
\label{noiseself}
\end{equation}

If the value of $H_0$ is precisely known, the above variances can be
directly calculated for a sample, and be compared with the
observed deviations.  But the value of $H_0$ is still controversial,
which leaves us the only option of investigating the relative
variation of Hubble flows within a sample.  In other words,
if the expansion rate of a sample with $N$ objects is $H_N$, and
the expansion rate of a subsample with $n$ ($< N$) objects is
$H_n$, a comparison can be made between the
variation $\delta H_{nN}/H_N=(H_n-H_N)/H_N$ and its theoretical
expectation without knowing the absolute value of $H_0$.
Under the condition that $H_N\approx H_0$
(i.e., $\langle (H_N-H_0)^2/H_0^2\rangle^{1/2}\ll 1$),
the variance of $\delta H_{nN}/H_N$ is
\begin{eqnarray}
&&\Bigl\langle\Bigl({\delta H_{nN}\over H_N}\Bigr)^2\Bigr\rangle\nonumber\\
&\approx&\Bigl\langle\Bigl({\delta H_n^{(v)}\over H_0}\Bigr)^2
+\Bigl({\delta H_N^{(v)}\over H_0}\Bigr)^2-2\Bigl({\delta H_n^{(v)}\over H_0}
 \Bigr)\Bigl({\delta H_N^{(v)}\over H_0}\Bigr)\Bigr\rangle\nonumber\\
&+&\Bigl\langle\Bigl({\delta H_n^{(\epsilon)}\over H_0}\Bigr)^2
+\Bigl({\delta H_N^{(\epsilon)}\over H_0}\Bigr)^2
-2\Bigl({\delta H_n^{(\epsilon)}\over H_0}\Bigr)
\Bigl({\delta H_N^{(\epsilon)}\over H_0}\Bigr)\Bigr\rangle,
\label{variation}
\end{eqnarray}
where variances of $\delta H_n$ and $\delta H_N$ are calculated
through eqs.~(\ref{variance1}) and~(\ref{noiseself}) using their respective
window functions $W^i_n({\bf k})$ and $W^i_N({\bf k})$. 
The cross correlation of the expected variations is
\begin{eqnarray}
&&\Bigl\langle\Bigl({\delta H_n^{(v)}\over H_0}
\Bigr)\Bigl({\delta H_N^{(v)}\over H_0}\Bigr)\Bigr\rangle\nonumber\\
&=&\Omega_0^{1.2}\int {\rm d}^3k\,
{\rm Re}\bigl[W^i_n({\bf k})\hat k_i{W^j_N}^*({\bf k})\hat k_j\bigr]
\,{P(k)\over k^2},
\end{eqnarray}
where Re[...] denotes the real part of the argument, and $*$ denotes
complex conjugation.  The correlation of the noises is
\begin{equation}
\Bigl\langle\Bigl({\delta H_n^{(\epsilon)}\over H_0}\Bigr)
\Bigl({\delta H_N^{(\epsilon)}\over H_0}\Bigr)\Bigr\rangle=
\Bigl\langle\Bigl({\delta H_N^{(\epsilon)}\over H_0}\Bigr)^2\Bigr\rangle.
\end{equation}
The size of the samples I discuss in this paper is sufficiently large
that their cosmic+sampling variance
$\langle (\delta H_N/H_0)^2\rangle^{1/2}$ is less than 3 percent
(Shi \& Turner 1998).  Eq.~(\ref{variation})
is therefore a very good approximation.
The calculation of the variation $\delta_{nN}$ and its variance,
on the other hand, has taken full account of both cosmic variance and
sampling variance.  They therefore do not depend on the sampling
size of the (sub)sample.  Of course for a small size subsample with
too few objects, the variance of $\delta H_{nN}$
becomes large, so that a comparison between observations and the
theoretical expectation will yield a less significant result. 
But the result is still statistically sound.

\begin{figure}
\psfig{figure=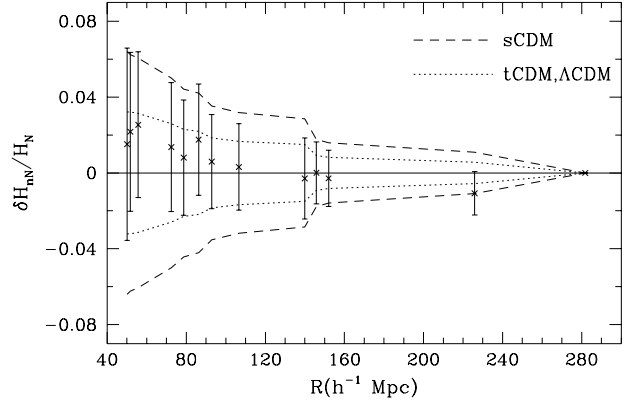,width=3.2in,angle=0}
 \caption{Hubble flow variations (with 1$\sigma$ errorbars)
vs. the depth of subsamples of the Type Ia SN
sample. Curves are noise-free 1$\sigma$ expectation from three models.}
\end{figure}

\begin{figure}
\psfig{figure=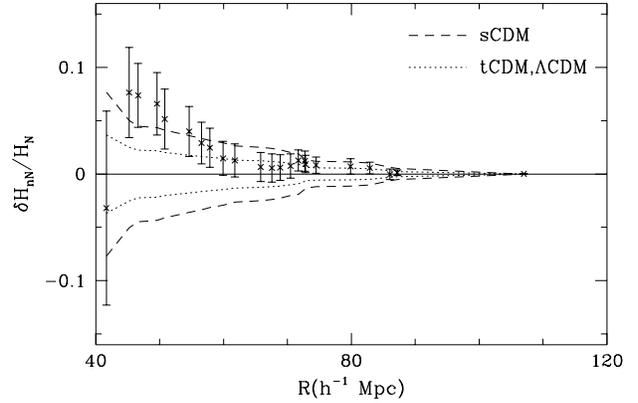,width=3.2in,angle=0}
 \caption{The same as fig. 1 but for the TF cluster sample.}
\end{figure}

In figures 1 and 2 I plot the Hubble flow variation $\delta H_{nN}$,
compared with the noise-free
model expectation of the standard deviation $\langle(\delta H_{nN}/H_N)^2
\rangle^{1/2}$, as a function of the maximal depth $R$ of subsamples
(defined to include $n$ most nearby objects),
for a sample of 20 Type Ia supernovae (SNe) (Riess et al. 1996)
and a sample of 36 clusters with Tully-Fisher (TF) distances
(Willick et al. 1997), respectively. 
Three representative models are chosen for comparison: (1)
the standard CDM model (sCDM), with $\Omega_0=1$, $h\equiv H_0/(100
{\rm km/sec/Mpc})=0.5$ and the power spectrum index $n_{\rm ps}=1$;
(2) the tilted CDM model (tCDM), with $\Omega_0=1$, $h=0.5$
and $n_{\rm ps}=0.7$; (3) the vacuum energy $\Lambda$ dominated
flat CDM model ($\Lambda$CDM),
with $\Lambda=0.7$, $\Omega_0=0.3$, $h=0.7$, and $n_{\rm ps}=1$.
The power spectra of these models
have the same functional form and {\sl COBE}-normalization as in
Shi (1997a).  The Type Ia SN sample does not show
significant detection of Hubble flow variations.  It therefore favors models
with smaller power on $\sim 40h^{-1}$ to 200$h^{-1}$ Mpc scales, because
otherwise its Hubble flow variation would be significantly larger.
The TF cluster sample shows a significant detection of Hubble
flow variations on 45$h^{-1}$ to 60$h^{-1}$ Mpc scale.
Its implication on models, however, is not obvious.

To quantify the implications of figs. 1 and 2,
a likelihood analysis is needed.  Since different subsamples are not
independent, we need to know the expected correlation
between the Hubble flow variations of two different subsamples,
$\delta H_{nN}$ and $\delta H_{mN}$ ($m\le n$).  This correlation is
\begin{eqnarray}
\Sigma_{nm}&=&\Bigl\langle\Bigl({\delta H_{nN}\over H_N}\Bigr)
                          \Bigl({\delta H_{mN}\over H_N}\Bigr)
\Bigr\rangle\nonumber\\
&\approx&\Omega_0^{1.2}\int {\rm d}^3k\,
{\rm Re}\Bigl\lbrace\Bigl[W^i_n({\bf k})-W^i_N({\bf k})\Bigr]\hat k_i\times
\nonumber\\
&&\Bigl[{W^j_m}^*({\bf k})-{W^j_N}^*({\bf k})\Bigr]\hat k_j\Bigr\rbrace
\, {P(k)\over k^2}\nonumber\\
&+&\Bigl\langle\Bigl({\delta H_n^{(\epsilon)}\over H_0}\Bigr)^2
-\Bigl({\delta H_N^{(\epsilon)}\over H_0}\Bigr)^2\Bigr\rangle.
\label{diff}
\end{eqnarray}
Given vector $(\Delta H)_n=\delta H_{nN}$ ($n=n_{\rm min}, n_{\rm min}+1,
..., N-1$) measured from a real sample $D$,
the likelihood of a cosmological model with a set of parameters $\theta$ is
\begin{equation}
P\langle \theta\vert DI\rangle= {1\over {\cal N}\vert\Sigma\vert^{1/2}}
\exp\Big[-{1\over 2}({\Delta H})^T(\Sigma)^{-1}(\Delta H)\Bigr],
\label{likelihood}
\end{equation}
with a normalization
\begin{equation}
{\cal N}=\int {\rm d}\theta {1\over\vert\Sigma\vert^{1/2}}
\exp\Big[-{1\over 2}({\Delta H})^T(\Sigma)^{-1}(\Delta H)\Bigr].
\label{norm}
\end{equation}
I test $\delta H_{nN}$ here instead of $H_n-H_{n-1}$ as
in Shi (1997a), because no advantage was found using the latter quantity.
The statistical tests of the two quantities, however,
are almost equivalent and give similar results.

\section{Results and Discussions}
I apply eqs.~(\ref{pechub}) to (\ref{norm}) to two sets of
{\sl COBE}-normalized CDM models, the
open CDM models ($\Omega_0<1$ and $\Lambda=0$) and the flat $\Lambda$CDM models
($\Omega_0+\Lambda=1$).  The parameters of the models are chosen to be
$\Omega_0$, $h$ and $n_{\rm ps}$.  For open (flat) CDM models, 
I choose the prior distribution of the parameters to be
$0.2<\Omega_0<1$ ($0.3<\Omega_0<1$), $0.5< h< 0.8$ and $0.7< n_{\rm ps}
< 1.2$, with uniformly distributed likelihood.
The ranges of parameters are chosen conservatively
to reflect measurements of mass densities on the galaxy cluster scale
(Carlberg, Yee and Ellingson 1997),
quasar lensing statistics in the case of flat CDM models (Kochanek 1996),
Cosmic Microwave Background
(Lineweaver and Barbosa 1997) and the Hubble constant (Freedman 1997).
The final likelihood distribution functions after taking into account
Hubble flow variations are then calculated according to
eq.~(\ref{likelihood}), with $\theta$ being a three dimensional
parameter space $(\Omega_0, h, n_{\rm ps})$.  Figures 3--6 show on slices
of constant $n_{\rm ps}$ the 68$\%$ and $95\%$ C.L. contours in
the 3-D parameter space, for open CDM models and flat CDM models, using
the Type Ia SN sample and TF cluster sample.  Only subsamples with a depth
of more than $40h^{-1}$ Mpc are used to ensure the validity of 
linear perturbation theory of gravity. 

Figures 3 to 6 clearly show that models with smaller powers are favored
by both samples.  Models with $\Omega_0\sim $0.3 to 0.4 fit data
quite well regardless of the choice of $h$, $n_{\rm ps}$ or the geometry
of the universe.
$\Omega_0=1$ models, on the other hand, are strongly disfavored unless
$n_{\rm ps}$ and $h$ approach the lower end of their allowed ranges.
The result diverges from a previous belief that
peculiar velocity fields favor a high matter density universe.
The figures also show that the Hubble variation method is a more
discriminative test to {\sl COBE}-normalized open models.

\begin{figure}
\psfig{figure=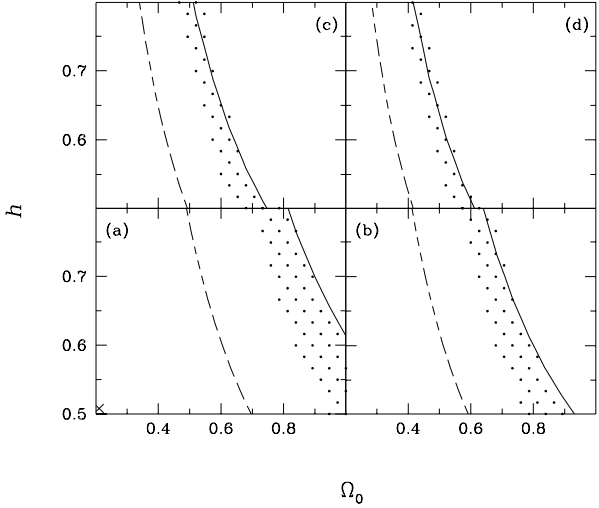,width=3.1in,angle=0}
 \caption{The 68$\%$ (the dashed line) and $95\%$ (the solid line)
C.L. contours in the $(\Omega_0,h,n_{\rm ps})$
parameter space on different $n_{\rm ps}={\rm constant}$ slices, for open CDM
models using the Type Ia SN sample.  Areas to the left of the contours are
allowed at their corresponding confidence levels.
Shaded regions are allowed at the 2$\sigma$ level
by the X-ray cluster temperature function.  The cross denotes the best fit.
(a) $n_{\rm ps}=0.725$;
(b) $n_{\rm ps}=0.875$;
(b) $n_{\rm ps}=1.025$;
(b) $n_{\rm ps}=1.175$.}
\end{figure}
\begin{figure}
\psfig{figure=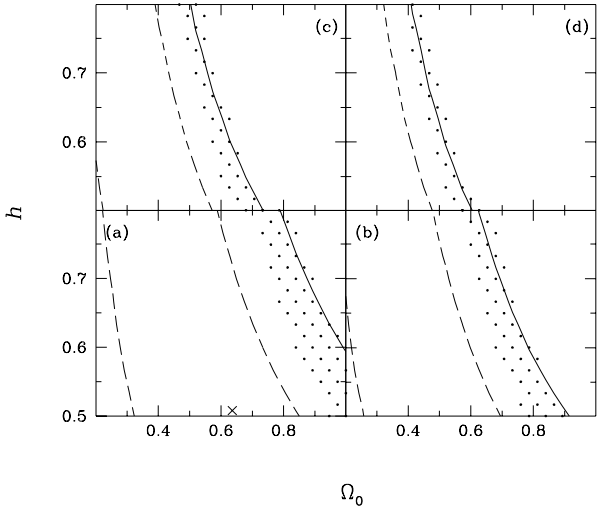,width=3.1in,angle=0}
 \caption{The same as fig. 3 but using the TF cluster sample.  Areas
sandwiched between two dashed lines are allowed at 68\% C.L.  Areas
to left of the solid line are allowed at 95\% C.L.}
\end{figure}
\begin{figure}
\psfig{figure=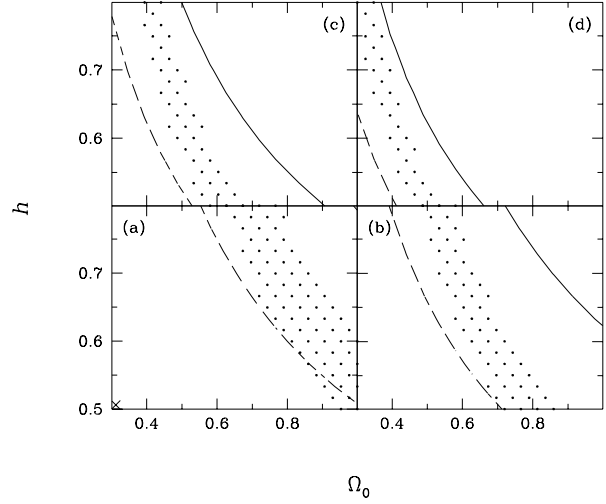,width=3.1in,angle=0}
 \caption{The same as fig. 3 but for flat $\Lambda$CDM models.}
\end{figure}
\begin{figure}
\psfig{figure=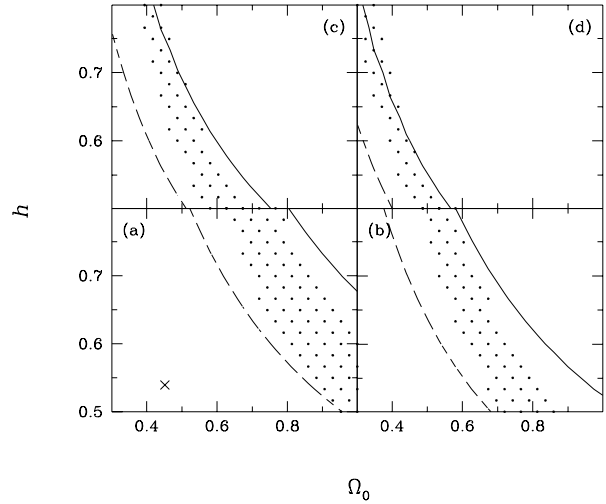,width=3.1in,angle=0}
 \caption{The same as fig. 3 but for flat $\Lambda$CDM models using
the TF cluster sample.}
\end{figure}

Also plotted in the figures is
the parameter space allowed by the X-ray cluster temperature function
constraint, $\sigma_8\Omega_0^\alpha=0.52\pm 0.04$ with
$\alpha =0.52-0.13\Omega_0$ for $\Lambda$CDM and $\alpha =0.46-0.10\Omega_0$
for open CDM (Eke et al. 1996; see also similar results of Viana and Liddle
1996, and Pen 1996), at the 2$\sigma$ level, if its error
is taken at a face value.  Since the X-ray cluster temperature function
probes the power spectrum on $\sim 8h^{-1}$ Mpc scale, while
Hubble flow variation calculation probes mainly scales from
$\sim 30h^{-1}$ to $\sim 200h^{-1}$ Mpc (as shown in figure 7),
their consistency is not automatically guaranteed.  Figures 3 to 6
show that the two methodologies do give consistent results for
individual samples.

\begin{figure}
\psfig{figure=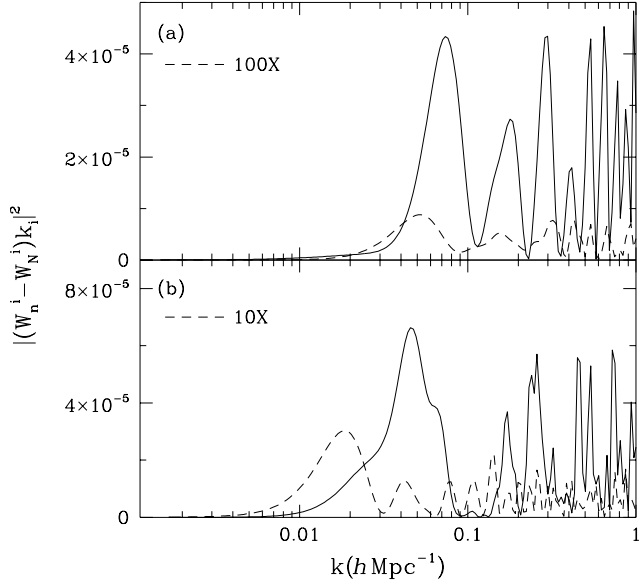,width=3.3in,angle=0}
 \caption{The window function vs. its scale.
(a) the TF cluster subsamples with $n=15$ (the solid line)
and $n=N-1=35$ (the dashed line, multiplied by 100);
(b) the Type Ia SN subsamples with $n=8$ (the solid line)
and $n=N-1=19$ (the dashed line, multiplied by 10).
Only one direction of ${\bf k}$ is shown.  The plotted function
looks similar in other ${\bf k}$ directions.}
\end{figure}

When both samples are used to calculate a joint likelihood of models,
models with smaller powers are more strongly favored, as seen in
figure 8 and 9.  However, while figure 9 shows that the Hubble flow
result is still compatible with the X-ray cluster constraints in the
{\sl COBE}-normalized $\Lambda$CDM models,
figure 8 indicates that the two are marginally {\sl incompatible}
in the {\sl COBE}-normalized open CDM models.
Does it hint that the flat universe with a
vacuum energy is a more likely scenario for our universe?  It is
probably premature to say so, because the inconsistency is only
marginal.  At this stage, it may tell us more about the smoother distribution
of the likelihood in $\Lambda$CDM models relative to open CDM
models, than about the favoring of one class of models
against the other.  If, however, the gap between the Hubble variation
result and the X-ray cluster result widens in open CDM models
in the future, we may need to seriously consider its cosmological
implications.

\begin{figure}
\psfig{figure=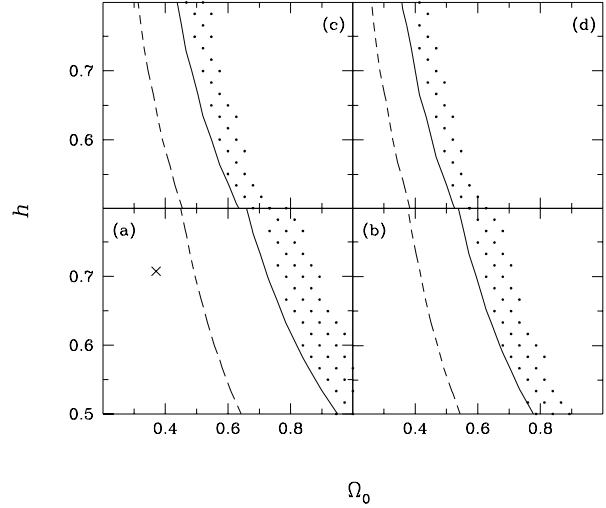,width=3.1in,angle=0}
 \caption{The joint likelihood distribution using both samples, for
open CDM models.  Legends are the same as in figure 3 to 6.}
\end{figure}

\begin{figure}
\psfig{figure=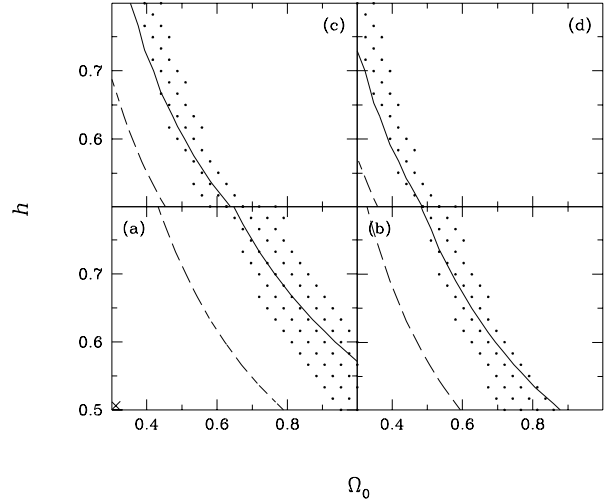,width=3.1in,angle=0}
 \caption{The joint likelihood distribution using both samples, for
flat $\Lambda$CDM models.  Legends are the same as in figure 3 to 6.}
\end{figure}

It is also interesting to compare the Hubble variation result to the
power spectrum inferred from the galaxy and cluster correlation analysis
of Peacock and Dodds (1994).  Since the two approaches both test
$\sim 100h^{-1}$ Mpc scales, certain consistency is expected.
Figure 10(a) plots $\Omega_0^{1.2}P(k)$
(which the Hubble flow variation truely measures)
of a number of open CDM models
along the 95$\%$ C.L. contour in fig. 8.  Since these models are
marginally allowed at 95$\%$ C.L., the upper envelope of their 
$\Omega_0^{1.2}P(k)$ curves in the tested range of $0.02\,{\rm Mpc}^{-1}
\la k/h\la 0.2\,{\rm Mpc}^{-1}$ represents a reasonable upper limit to
$\Omega_0^{1.2}P(k)$ of the {\sl COBE}-normalized open CDM models.
A similar limit can also be obtained for the {\sl COBE}-normalized
$\Lambda$CDM models. Figure 10(b)
shows that this upper limit for the {\sl COBE}-normalized
open CDM models is lower than the $\Omega_0^{1.2}P(k)$ of
Peacock and Dodds (1994), had our universe had $\Omega_0$ approaching 1.
This upper limit is consistent with Peacock and Dodds'
result only if the universe has $\Omega_0\la 0.6$,
given the $\Omega_0^{-0.3}$ dependence of their deduced $P(k)$.
For $\Lambda$CDM models, the upper limit on $\Omega_0^{1.2}P(k)$ is
consistent with the result of Peacock and Dodds even if $\Omega_0=1$,
although the consistency at $\Omega_0=1$ is marginal.
The comparison once again shows that small $\Omega_0$ is favored, and
that the test of Hubble flow variation is more discriminative to
the {\sl COBE}-normalized open CDM models than to
the {\sl COBE}-normalized $\Lambda$CDM models.

One major shortcoming of the Hubble flow variation test is that the likelihood
distribution is not symmetric relative to the peak probability (figs. 3 to 6
and figs. 8 and 9)
so that it has less power discriminating against models predicting too small
Hubble flow variations than against models with too large Hubble flow
variations.  This is due to the fact that the noise term in eq.~(\ref{diff})
dominates the Hubble flow variation when the true density fluctuations are
small.  Thus the key to increase the testing power on
models with small $P(k)$ (and on models with larger $P(k)$ to a lesser
degree), is to reduce the noise term with better
distance measurements, and more objects in the same sample volume.
Since the observed and expected $\delta H_{nN}/H_N$ is typically a
few percent at a depth of $\la 5,000$ to 10,000 km/sec,
the noise contribution to $\delta H_{nN}/H_N$
has to be $\la 1\%$ to ensure a significant detection of
$\delta H_{nN}/H_N$ and a little skewed likelihood distribution.
For Type Ia SN samples, where distance measurement errors are
typically 5$\%$ and random motions due to local non-linearities
contribute $\sim 5\%$ of recession velocities at $\la 10,000$ km/sec,
the number of Type Ia SNe has
to be $\ga 50$ within $\sim 10,000$ km/sec to reduce the noise term to
$\la 1\%$.  Apparently the sample used here (with the number
around 15) is not enough 
(as shown by the one-sided likelihood distribution in figs. 3 and 5).
But as the number of Type Ia SNe observed increases rapidly (for instance,
there are 25 Type Ia SNe below 10,000 km/sec in an unpublished
data set of Riess), it is hopeful that within the next several years a lot
more precise limit can be put on the power spectrum from both
the high end and the low end.

For TF cluster samples, the noise contribution is dominated by distance
measurement errors ($\approx 15\%$).  Therefore, there has to be about
200 clusters within 10,000 km/sec to significantly boost the power of the
Hubble flow variation method.  This is not easy but still hopeful with
larger and deeper surveys of the sky, and with
distance measurements of clusters refined to better than about $10\%$.

\begin{figure}
\psfig{figure=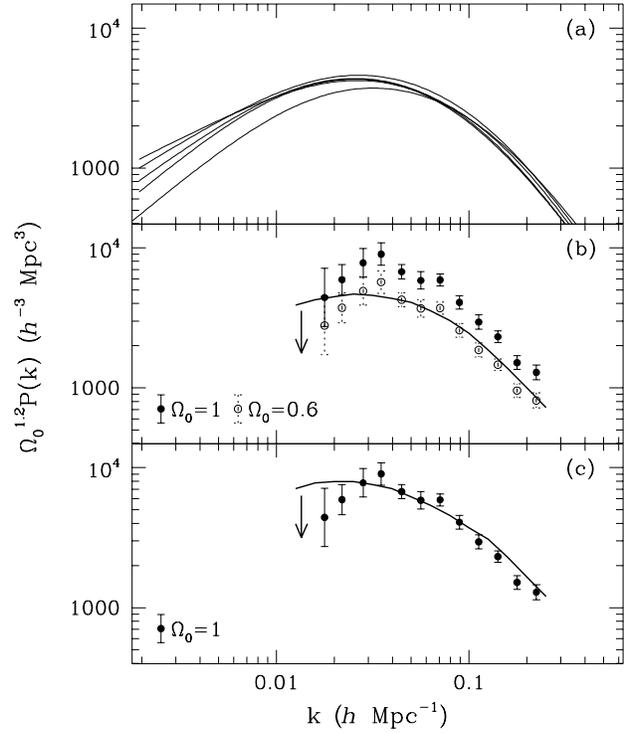,width=3.2in,angle=0}
 \caption{
Panel (a) plots $\Omega_0^{1.2}P(k)$
of open CDM models that lie on the 95\% C.L. contours of figure 8.
In (b) and (c), solid curves are the upper limits
on $\Omega_0^{1.2}P(k)$ from the Hubble flow variation test (see text)
for the open CDM models (panel (b)) and
the flat $\Lambda$CDM models (panel (c)).  For comparison are
the results from Peacock and
Dodds (1994) for $\Omega_0=1$ (solid data points)
and $\Omega_0=0.6$ (open data points).}
\end{figure}

\section{Acknowledgments}

The author thanks Aspen Center for Physics, where many of the issues
discussed here were raised, for its hospitality.  Thank is also due
to the referee for providing valuable suggestions and criticisms.
The work is supported by grants NASA NAG5-3062 and NSF PHY95-03384 at UCSD.

\end{document}